\documentstyle[12pt,psfig]{article}
\def\gsim{\mathrel{\hbox{\rlap{\lower.55ex \hbox {$\sim$}}
                   \kern-.3em \raise.4ex \hbox{$>$}}}}

\def\rerg{\rm erg}
\def\rs{\rm s}
\def\rs1{\rm s^{-1}}

\def\rcm{\rm cm}
\def\rcm2{\rm cm^{-2}}

\def\flux{\rerg\ \rcm2\ \rs1}

\def\c2r{\chi^2_\nu}
\def\c2{\chi^2}

\def\etal{{\it et al. }}
\def\apj{{\it Astrophys. J. }}
\def\apjs{{\it Astrophys. J. Suppl. Ser. }}
\def\pasj{{\it Publs. Astron. Soc. Japan }}

\def\aa{{\it Astron. Astrophys. }}
\def\aas{{\it Astron. Astrophys. Suppl. Ser. }}

\begin{document}

\vspace{0.4cm}

{\bf Published in Science, 290, 953 (November, 3 2000)}
\vspace{0.5cm}
\begin{center}
{\LARGE \bf Discovery of a transient absorption  edge in the
X--ray spectrum of GRB 990705}

\vspace{0.8cm}

{ \bf Lorenzo Amati$^{1\ast}$, 
Filippo Frontera$^{1,2}$,
Mario Vietri$^{3}$,
Jean J.M. in 't Zand$^{4}$,
Paolo Soffitta$^{5}$,
Enrico Costa$^{5}$, 
Stefano Del Sordo$^{6}$,
Elena Pian$^{1}$, 
Luigi Piro$^{5}$,
Lucio A. Antonelli$^{7}$,
D. Dal Fiume$^{1}$, 
Marco Feroci$^{5}$, 
Giangiacomo Gandolfi$^{5}$, 
Cristiano Guidorzi$^{2}$, 
John Heise$^{4}$,
Erik Kuulkers$^{4}$,
Nicola Masetti$^{1}$, 
Enrico Montanari$^{2}$, 
Luciano Nicastro$^{6}$,
Mauro Orlandini$^{1}$,
Eliana Palazzi$^{1}$ 
}

\end{center}

\vspace{0.7cm}

{ \noindent $^{1}$ Istituto di Tecnologie e Studio delle Radiazioni Extraterrestri, CNR., Via Gobetti 101, 
40129 Bologna, 
Italy
\newline $^2$ Dip. Fisica, Universita' di Ferrara, Via Paradiso 12, 44100 
Ferrara, Italy 
\newline $^{3}$ Dip. Fisica, Terza Universit\'a di Roma, Via della Vasca Navale 84, 00146
Roma, Italy
\newline $^{4}$ Space Research Organization Netherlands, 
Sorbonnelaan 2, 3584 CA Utrecht, The Netherlands 
\newline $^{5}$ Istituto di Astrofisica Spaziale, CNR., Via Fosso del 
Cavaliere, 00133 Roma, Italy
\newline $^{6}$ Istituto di Fisica Cosmica con Applicazioni all' Informatica, 
Via U. La Malfa, 153, 90146 Palermo, Italy
\newline $^{7}$ Osservatorio Astronomico di Roma, Via Frascati 33, 00040 Monte 
Porzio Catone (RM), Italy
\newline
\newline $^{\ast}$ To whom correspondence should be addressed E--mail: amati@tesre.bo.cnr.it
}

\vspace{1.0cm}
%\hrule

\vspace{1.0cm}

{\small \bf
We report  the discovery of a transient equivalent hydrogen column density with
an absorption edge at $\sim$3.8 kiloelectron volts in the spectrum of the prompt x--ray 
emission of gamma--ray burst (GRB) 990705. This feature can be satisfactorily modeled with a 
photo-electric absorption by a medium located at a redshift of 
$\sim$0.86 and with an iron abundance of $\sim$75 times the solar one. 
The transient behavior is attributed to
the strong ionization produced in the circumburst medium by the GRB photons. 
The high iron abundance found
points to the existence of a burst environment enriched by a supernova along
the line of sight. The supernova explosion is estimated to have occurred
about 10 years before the burst. Our results agree with 
models in which GRBs originate from the collapse of very massive stars and 
are preceded by a supernova event.}

\vspace{0.5cm}

The nature of the progenitors of celestial GRBs
is an open issue of key astrophysical importance. Collapse of
massive fast rotating stars $-$ the hypernova model ({\it 1}) $-$ or delayed
collapse of a rotationally stabilized neutron star $-$ the supranova model ({\it 2}) $-$
are among the favored scenarios for the origin of these events.
Both models predict that the pre-burst environment is characterized by a high
gas density, either as a result of strong winds from the massive
progenitor in the case of a hypernova or because of a supernova (SN) event in the case 
of the supranova model ({\it 3}). In the latter case the environment is expected
to be enriched in heavy elements. The progressive
photoionization of the circumburst material (CBM)  by the GRB photons should
produce in the burst x--ray spectrum  transient low--energy cut-off and
K--edge absorption features  of the elements in the CBM ({\it 4}).
The detection of such features may allow us to estimate the
density and composition of the CBM, the GRB redshift, 
and, ultimately, the nature of the GRB progenitor. \\
Owing to the coalignment of two detection units of the gamma ray burst
monitor (GRBM, 40 to 700 keV) ({\it 5,6})
with the two wide field cameras (WFCs, 2 to 26 keV) ({\it 7}),
the Italian--Dutch x--ray mission BeppoSAX can provide not only  
arc minute localizations of GRBs, but also measurements of their spectra
in a broad (2 to 700~keV) energy band ({\it 8}).
Among the GRBs detected by the BeppoSAX WFC and GRBM, the event of
5 July 1999 (GRB 990705) is the second brightest in $\gamma$--rays (40--700~keV)
after GRB 990123 and ranks in the top 15\% in x--rays (2--26~keV). 
This burst
triggered the GRBM on 5 July at 16:01:25 universal time and was positioned with an
error radius of 3 arc min at right ascension $\alpha(2000) = 05^{h}
09^{m} 52^{s}$ and declination $\delta(2000) = -72^{\circ} 08' 02''$, in a direction
close to the edge of the Large Magellanic Cloud. 
Optical and near--infrared observations of the GRB 990705 location
led to the discovery of a reddened fading counterpart
and a possible host galaxy ({\it 9}).
Recently, Holland et al. ({\it 10}) have imaged the GRB 990705 field with
the Hubble Space Telescope, detecting a spiral galaxy at
the position of the GRB. Although the distance of this galaxy is not known,
its size and brightness are compatible with a redshift $z \le 1$ ({\it 11}).

A 120,000 s follow--up observation with the BeppoSAX narrow-field
instruments ({\it 12}) was also performed starting $\sim$11 hours after the GRBM 
trigger. In the first 7 hours of observing time, an x--ray  source
of 3.3$\sigma$ significance, corresponding to 1.9 ($\pm$0.6) $\times$10$^{-13}$
$\flux$,
was detected from a 2 arc min radius region centered on the near--infrared
counterpart position. The source, 1SAX J0509.9-7207, was not visible
after this time. On the basis of the fading behavior, 1SAX J0509.9-7207 is
the most likely candidate of the GRB 990705 x--ray afterglow.
\vspace{0.3cm}

The event exhibits a highly structured pulse, a duration of 42~s in
$\gamma$--rays and a longer duration ($\sim$60~s) in x--rays (Fig. 1).
Following the investigation results obtained with a sample of BeppoSAX
GRBs ({\it 8}), we observed the spectral evolution
of the GRB prompt emission by accumulating WFC and GRBM spectra in 7
adjacent time  intervals (Fig. 1). 
The 2 to 700 keV spectra of time intervals from C to G can be fit with
a simple power law model [$I(E)\propto E^{-\Gamma}$] with a continuously
variable photon index in the range from $1.22\pm0.02$ to $2.24\pm0.28$
(the uncertainties quoted hereafter are 1 standard deviation).
This hard--to--soft evolution is typical of GRB prompt emission
spectra({\it 8}).
Instead, the spectra of the first two time intervals (A and B in Fig.~1),
of 6~s and 7~s duration, respectively, cannot be described either
by this model (see Fig.~2) or by the smoothed broken power--law model
proposed by Band et al. ({\it 13}) for GRB spectra.
The description with a power--law photoelectrically absorbed by
a gas with cosmic abundance ({\it 14}) within our galaxy ($z = 0$) provides a 
good fit ($\chi^2/dof = 5.55/9$) for the slice A
with equivalent hydrogen column density 
$N_{\rm H} = 8.7 (\pm3.6) \times 10^{22}$~cm$^{-2}$ and power--law
photon index $\Gamma = 1.08 \pm 0.03$. However for
time slice B the same model does not  provide a good description
($\chi^2/dof = 29.9/9$); the depression between 4 and 6 keV, which is
apparent in this spectrum (Fig.~2, middle panel), is still there.
A possible description of the feature ($\chi^2/dof = 5.8/7$) is
obtained by adding to the photoelectrically absorbed power--law, 
whose best fit parameter values are 
$N_{\rm H} = 3.5 (\pm 1.4) \times 10^{22}$~cm$^{-2}$ and $\Gamma = 1.08\pm 0.02$, 
an absorption edge of energy E$_{edge}$=3.8$\pm$0.3~keV, and optical
depth $\tau$=1.4$\pm$0.4. More naturally, the best fit of the time slice B
spectrum ($\chi^2/dof = 5.5/7$) is obtained with a
photoelectrically absorbed power law with iron relative
abundance with respect to the solar ({\it 15}), Fe/Fe$_\odot$, and z of
the absorbing medium left free to vary. In  Fig.~3 we show the
result; we obtain a photon index $\Gamma = 1.09 \pm 0.02$, an absorption column
density of $N_{\rm H} = 1.32 (\pm 0.30) \times 10^{22}$~cm$^{-2}$, a relative
abundance Fe/Fe$_\odot$=75$\pm$19 and a redshift $z = 0.86\pm 0.17$.
The value of the photon index agrees with the spectral hardness
expected at the early times of the event.
We have also tested other elements such as Ca, Cr, Co, and Ni with K--edge close to or
higher than 3.8 keV. With Ca we still obtain a satisfactory fit
but the abundance required (Ca/Ca$_\odot = 1083 \pm 285$) is
higher than the values found in Ca--rich astrophysical media ({\it 16}).

We investigated whether the best model found for the B spectrum can also give a good
description of the spectrum measured during the time interval A. The result is positive,
but the statistics of the data do not allow us to constrain all the model
parameters. Assuming the best fit values of Fe/Fe$_\odot$, z and $N_{\rm H}$ found for the
time interval B, the A spectrum is equally well described ($\chi^2/dof = 8/10$) by this
model with  best fit photon index $\Gamma = 1.09 \pm 0.02$, coincident  with that found 
for the time interval B. Leaving only Fe/Fe$_\odot$ free to vary, the best fit 
($\chi^2/dof = 7.9/10$) is found for  Fe/Fe$_\odot = 71 \pm 22$. We can conclude 
the best fit model parameters found for the B spectrum also give a good description of the
data during the first time interval.

\vspace{0.3cm}

The total 2--700~keV fluence of GRB 990705 calculated with the time averaged spectrum
is $9.3 (\pm 0.2) \times 10^{-5}$~erg cm$^{-2}$.
Assuming the best fit value of $z$, a standard cosmology
with $H_{\rm 0} = 70$~km s$^{-1}$ Mpc$^{-1}$ and q$_{0} = 0.5$ and isotropic
emission, it corresponds to a released energy of $\sim 1.2 \times
10^{53}$~erg.
For comparison, in the time intervals A+B, the 2 to 700 keV fluence is 
$\sim 3.8\times 10^{-5}$~erg cm$^{-2}$, corresponding to a released energy 
of $\sim 4.7 \times 10^{52}$~erg, with almost all of it  
emitted above the energy of the absorption feature (3.8 keV).

The iron abundance inferred above points to
the existence of an environment iron--rich for a SN along
the line of sight. The most likely location for this material is
in the immediate surroundings of the burst (see below), and we 
shall neglect the possibility of a chance alignment along the 
line of sight. Metal enrichments by a factor of about 100 
above the solar value are considered typical of SNe
(and of no other astrophysical environment). This is true 
for type I SNe, and for type II SNe with massive 
progenitors, especially in the case of strong mass loss during 
the main--sequence phase ({\it 17}).
Assuming the SN progenitor to have a mass $M_{pr} = 10 m_1 M_\odot$, 
and the supernova remnant (SNR) to be distributed in a spherical
shell around the origin of
the GRB site, we can estimate the ejecta distance $D$ from the 
burst location from $ M_{pr}/m_p = 4\pi D^2 N_H$ (where $m_p$ is the 
proton mass). For the value $N_{\rm H} = 1.3\times 10^{22}$ cm$^{-2}$ derived above,
we get $D \approx 3\times 10^{17} d_{17} m_1^{1/2}$~cm, where $d_{17}$ is the distance in 
units of $3 \times 10^{17}$~cm. This
distance is actually an upper limit. Should the SNR be highly ionized even
before the burst, $N_{\rm H}$, which only measures the neutral column depth, 
would be a poor approximation for the total hydrogen column depth,
pushing $D$ to smaller values. 
For this value of $D$, one can estimate the time elapsed between the SN
explosion and the burst. Assuming a SN ejecta speed of
$10^4 v_4$~km s$^{-1}$,
we get $\delta\!t \sim 10 d_{17}/v_4$~years. But the major test
comes from showing that the absorption edge must disappear within a few seconds
after the burst onset. To show that this is indeed the case, we notice
first that
the large optical depths $\tau \approx 1.4$ inferred from our observations 
imply that most photons above the ionization threshold will be captured
as they fly through the SNR, until near--complete photoionization is
achieved. The required number of photons for the complete photoionization 
of iron can be estimated, from the above parameters, as $\approx 5 \times 
10^{56} m_1$, taking into account Auger ionization ({\it 4}) and including the exceptional 
iron abundance detected here, $75$ times higher than the solar value. On the 
other hand, during the first 13~s (duration of the time intervals A+B), this
burst
emits a time--averaged photon spectrum $d\!n = 4.2 d\!E/E$ photons cm$^{-2}$
s$^{-1}$ in the observer's frame. Within this time interval, and for a 
standard cosmology, this corresponds to a total of $3.6\times 10^{57}$ 
ionizing photons, where every photon above the threshold has been weighted with 
the ratio of its cross--section to that at the threshold. Furthermore, 
the density due to a $10 m_1 M_\odot$ star dispersed in a volume of
radius $D$ is about $10^5$~cm$^{-3}$, implying long recombination 
time scales ($> 10^5$ s), depending on the exact, but rapidly varying, 
temperature. Thus, we conclude that the burst has 
about the right number of photons to cause the complete photoionization
of iron within the B time interval, with recombination providing no real
counter--effect. A similar conclusion, with a more elaborate computation,
has been reached, for generic parameters, by B\"ottcher \etal ({\it 4}).

The implication of the above discussion is that the iron--rich material is, 
most likely, located around the burst site and cannot be located by chance along 
the line of sight. In fact, from the column density and iron abundance derived
from the B spectrum, we have found that the absorbing material has an iron content 
$\sim 75$ times the solar value, and we have deduced that only 
a SNR can be responsible for this absorption. Calling $R_{SNR}$ the radius of the SNR, 
and $D$ its distance from the burst site, if the SNR were really located by
chance along the line of sight, it would intercept  only a fraction
$\delta\!\Omega \sim (R_{SNR}/D)^2$ of all burst photons. For a chance 
alignment, $R_{SNR} \ll D$, so that $\delta\!\Omega \ll 1$; however, this
reduced number of burst photons would still have to ionize the whole SNR, i.e.,
several solar masses of matter. Because we have determined above that this is just 
about feasible for $D \approx 10^{17}\; cm$, it follows that this cannot be
accomplished for a chance alignment, where of course $D \gg 10^{17}\; cm$, 
and we should be able to see the absorption edge through the entire burst 
duration. Given that this is not the case, we deduce that the SNR cannot be located
by chance along the line of sight.

Lastly, the nondetection of an iron line (3$\sigma$ upper limit of
1.5 photons cm$^{-2}$ s$^{-1}$) from the SNR is consistent with our model.
Indeed, at least three factors may contribute to
making the iron line weak. First, the burst may be beamed, whereas the line 
reemission certainly is not.
Second, although we see the whole SNR, yet 
line reemission 
is diluted with respect to the burst duration T$_B$ by the light transit 
time $D/c$ in the ratio cT$_B$/D. Third, although fluorescence may be fast, 
recombination must be slow, owing to the low overall material densities on the order of
$10^5$ atoms/cm$^3$. Still, 
such a line should be present at later times, when the afterglow 
decreases sufficiently; four cases of this kind
have been reported [GRB970508 ({\it 18}), GRB970828 ({\it 19}), GRB991216 ({\it 20}),
GRB000214 ({\it 21})].

Our results favor models
in which GRBs originate from the collapse of very massive stars and are
preceded by a supernova--like explosion.

\vspace{0.5cm}

%\hrule

\vspace{0.5cm}

{\bf References and Notes}
\vspace{0.1cm}
{\noindent

1. B. Paczynski,
\apj {\bf 494}, L45 (1998).
 
2. M. Vietri and L. Stella,
\apj {\bf 507}, L45 (1998).
 
3. C. Weth, P. Meszaros, T. Kallman, M. Rees, 
\apj {\bf 534}, 581 (2000).

4. M. B\"ottcher, D.D. Dermer, W. Crider, E.P. Liang, 
\aa {\bf 343}, 111 (1999).

5. F. Frontera et al.,
\aas {\bf 122}, 357 (1997).

6. E. Costa et al., 
{\it Adv.Sp.Res.} {\bf 22/7}, 1129 (1998).

7. R.Jager et~al., 
\aas {\bf 125}, 557 (1997).

8. F. Frontera et~al., 
\apjs {\bf 127}, 59 (2000).

9. N. Masetti et~al., 
\aa {\bf 354}, 473 (2000).

10. S. Holland et al., {\it GCN Circular} 753 (2000).

11. J.G. Cohen et al.,
\apj {\bf  538}, 29 (2000).

12. G. Boella et~al., 
\aas {\bf 122}, 299 (1997).

13. D. Band et al., \apj {\bf 413}, 281 (1993).

14. R. Morrison, D. McCammon,
\apj {\bf 270}, 119. (1983).

15. E. Anders, N. Gredesse, 
{\it Geochim. et Cosmochim. Acta}, {\bf 53}, 197  (1989).

16. H. Tsunemi, E. Miyata, B. Aschenbach, J. Hiraga, D. Akutsu, 
\pasj, in press.

17. S.E. Woosley and T.A. Weaver,
{\it Ann. Rev. of Astron. and Astrophys.} {\bf 24}, 205 (1986).

18. L. Piro et~al.,
\apj {\bf 514}, L77  (1999).

19. A. Yoshida et~al.,
\aas {\bf 138}, 433 (1999)

20. L. Piro et ~al.,
{\it Science}, submitted

21. L.A. Antonelli et~al.,
\apj, submitted.

22. L. Amati et~al., 
\aas {\bf 138}, 403 (1999).

23. This research was supported by the Italian Space Agency (ASI) and Italian
National Research Council (CNR). Beppo--SAX is a joint program of ASI and
the Netherlands Agency for Aerospace Programs.
}

\clearpage

{\bf Fig. 1.} \\
{(\bf Top)} WFC light curve (2 to 28 keV) with 0.5~s time
resolution. {(\bf Bottom)} GRBM light curve (40 to 700 keV)
with 0.128~s time resolution. The vertical dotted lines limit the 7
intervals in which spectral analysis was performed. The typical error
is shown for each light curve on the left of the panel.

\vspace{1.2cm}

{\bf Fig. 2.} \\
Distribution of the residuals of the count spectra from the best--fit 
power--law model in the time intervals {\bf A (top)},
{\bf B} (middle), and {\bf C} (bottom).
The major deviations of the data from the model are apparent in the lowest
energy bin for the A spectrum and in the 3 to 6 keV band for the B spectrum,
whereas no statistically significant deviations from a power--law are
apparent for the C spectrum. The probability that 
the observed deviations are due to chance is 2.3$\times 10^{-2}$ for time
interval A and 9.0$\times 10^{-6}$ for time slice B. The higher chance probability obtained
for the deviations in the time interval A  is likely due to the lower statistical quality of 
these data (see text).
The possibility that instrumental effects caused the depression in B spectrum were
investigated with  negative results. The cross--calibration between
the two instruments
was verified by measurements of the Crab Nebula spectrum ({\it 22}).
The depression is also found in the ratio between the
GRB count rate spectrum of the time interval B and the Crab Nebula spectrum
taken with the same instrumentation. The depression is not apparent in the
spectra accumulated over the successive time intervals (see bottom panel
for the time slice C).

\vspace{1.2cm}

{\bf Fig. 3.} \\
Photon spectrum in the time slice B. The continuous plot shows the best--fit
curve obtained with a power--law plus a photoelectric absorption by a medium
at redshift $z = 0.86$,  column density N$_H$=1.3$\times$10$^{22}$
cm$^{-2}$, and iron abundance 75 times that of the sun.

\clearpage
%
%  Figure 1
%
\begin{figure}[!t]
\vspace{6.5cm}
\centerline{\psfig{figure=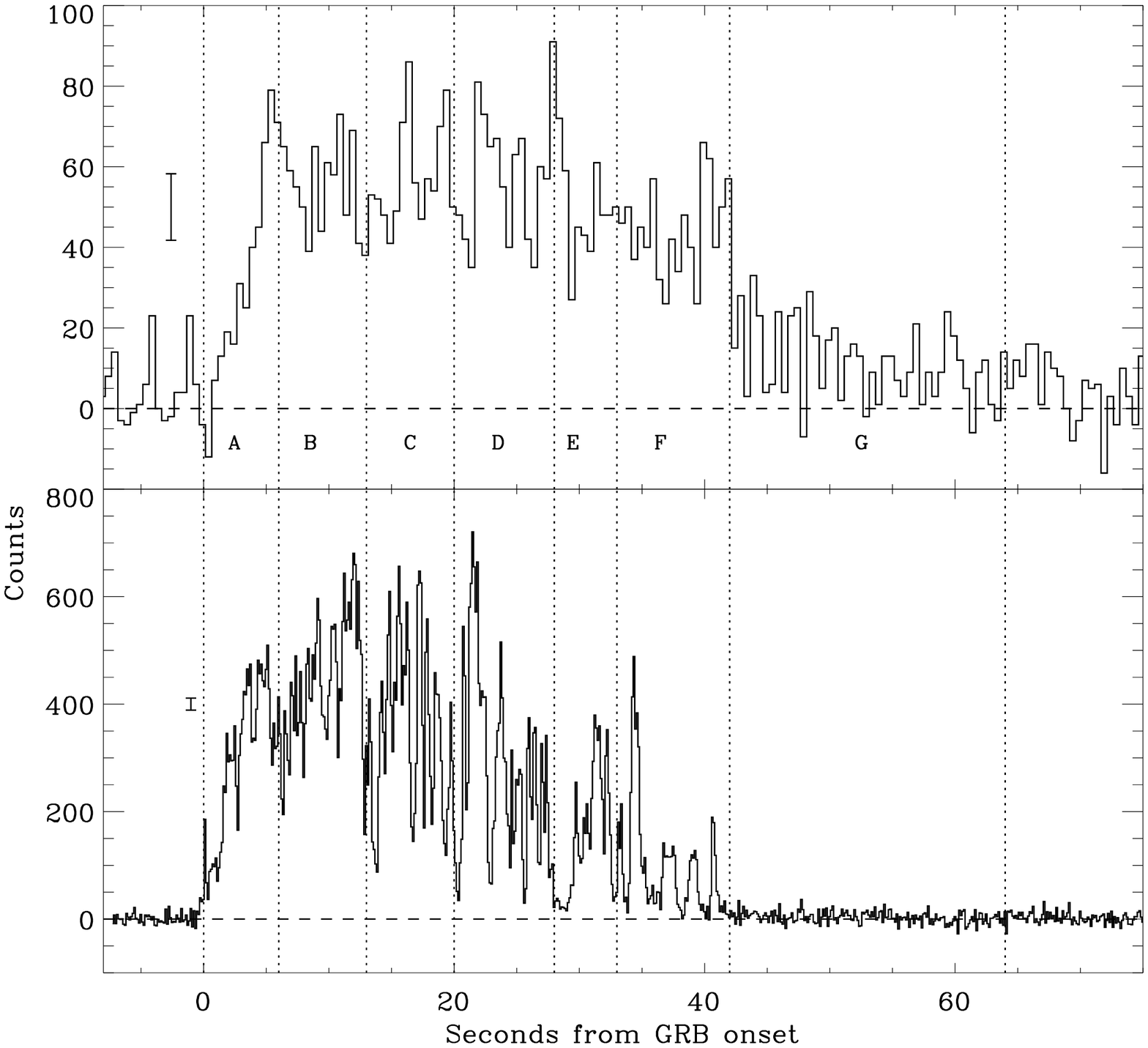,width=14cm}}
\vspace{2.0cm}
\end{figure} 

\vspace{1.5cm}
\centerline{\bf Fig. 1}

\clearpage
\thispagestyle{empty}

%
%Figure 2
%
\begin{figure}[!t]
\vspace{6.5cm}
\centerline{\psfig{figure=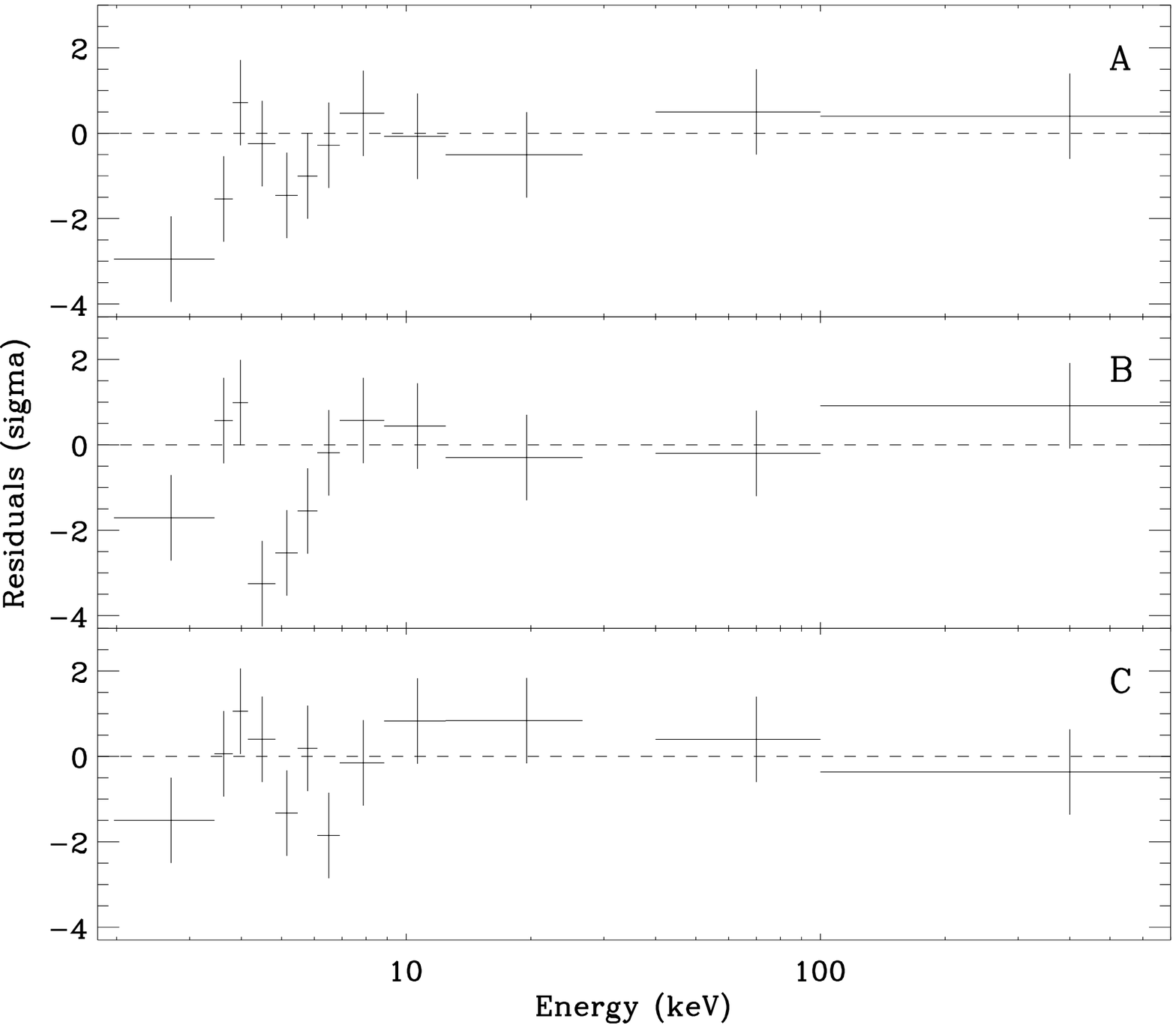,width=14cm}}
\vspace{2.0cm}
\end{figure} 

\centerline{\bf Fig. 2}

\clearpage
\thispagestyle{empty}
%
%  Figure 1
%
\begin{figure}[!t]
\vspace{6.5cm}
\centerline{\psfig{figure=zvphabs_fot.ps,width=14cm,angle=-90}}
\vspace{2.0cm}
\end{figure} 

\vspace{1.5cm}
\centerline{\bf Fig. 3}

\end{document}